\documentstyle[12pt,psfig]{article}

\newcommand{\beq}{\begin{equation}}
\newcommand{\eeq}{\end{equation}}
\newcommand{\beqa}{\begin{eqnarray}}
\newcommand{\eeqa}{\end{eqnarray}}

\newcommand{\slas}[1]{{#1}\!\!\!/}
\newcommand{\etal}{{\em et al.}}

\newcommand{\gsim}{\buildrel > \over {_\sim}}

\newcommand{\PRD}[3]{Phys.\ Rev.\ D {\bf {#1}}, {#2} ({#3})}
\newcommand{\PRL}[3]{Phys.\ Rev.\ Lett.\ {\bf {#1}}, {#2} ({#3})}
\newcommand{\NPA}[3]{Nucl.\ Phys.\ A {\bf {#1}}, {#2} ({#3})}
\newcommand{\NPB}[3]{Nucl.\ Phys.\ B {\bf {#1}}, {#2} ({#3})}
\newcommand{\PLB}[3]{Phys.\ Lett.\ B {\bf {#1}}, {#2} ({#3})}
\newcommand{\ZPC}[3]{Z. Phys.\ C {\bf {#1}}, {#2} ({#3})}

\begin{document}

\begin{titlepage}
\renewcommand{\thefootnote}{\fnsymbol{footnote}}
\makebox[2cm]{}\\[-1in]
\begin{flushright}
TUM/T39-98-12
\end{flushright}
\vskip0.4cm
\begin{center}
  {\Large\bf
    \setcounter{footnote}{1}
    Hard exclusive $J/\psi$ leptoproduction on polarized
    targets\footnote{Work supported in part by BMBF}
  }\\ 

\vspace{2cm}

\setcounter{footnote}{0}
M. V\"anttinen\footnote{Alexander von Humboldt fellow}
\setcounter{footnote}{6}
and L. Mankiewicz\footnote{On leave of absence from N. Copernicus
Astronomical Center, Polish Academy of Science, ul.\ Bartycka 18,
PL--00-716 Warsaw (Poland)}

\vspace{1 cm}

{\em  Physik Department, Technische Universit\"{a}t M\"{u}nchen, \\
D-85747 Garching, Germany}

\vspace{1cm}

{\em \today}

\vspace{1cm}

{\bf Abstract:\\[5pt]} \parbox[t]{\textwidth}{We consider
the exclusive production of $J/\psi$ mesons in 
polarized virtual-photon--proton collisions. We derive helicity
amplitudes for the dominant subprocess $\gamma^* g \rightarrow J/\psi \; g$
and show that polarization asymmetries vanish in the case of
collinear scattering. Thus, contrary to what has been suggested earlier,
this process is not a good probe of the polarized gluon distribution
$\Delta g(x,Q^2)$ of the proton.}

\end{center}
\end{titlepage}

\newpage
\renewcommand{\thefootnote}{\arabic{footnote}}
\setcounter{footnote}{0}


\setlength{\baselineskip}{8mm}

\section{Introduction}

Considerable attention has recently been devoted to study exclusive
meson production in QCD. In particular, there has been renewed interest
\cite{Ji-PRL78,Radyushkin} in the concept of nonforward
(also called off-forward, off-diagonal or asymmetric) parton
distributions of nucleons. The nonforward
distributions are defined as matrix elements of non-local twist-2 QCD
string operators
between nucleon states of unequal momenta. Thus, they present
a generalization of the concept of ordinary parton distributions on
one hand and of nucleon form factors on the other hand.

In physical terms, leading-twist (twist-2) nonforward parton
distributions characterize a process where one 
parton is extracted from a nucleon and another is returned,
carrying a different light-cone momentum fraction. Such processes
can be probed in exclusive reactions like the (virtual)
photoproduction of mesons \cite{Radyushkin,Brodsky,Hoodbhoy,Collins,MPW97}
and deeply virtual Compton scattering \cite{Ji-PRD55}.
There is currently significant experimental activity on these
topics at both collider \cite{HERA} and fixed-target
\cite{TJNAF,HERMES,COMPASS} energies.

The nonforward distributions become equal to their forward
counterparts only in the exact forward limit i.e., when the
difference between initial and final nucleon momenta vanishes.
On the other hand it has been argued by many authors
\cite{MPW97,Frankfurt,Martin97} that nonforward distributions which
determine the amplitude for exclusive meson production in the hard
diffractive limit have shape similar to corresponding forward
distributions. 

Because the probability amplitudes of exclusive reactions are
proportional to nonforward distributions, exclusive cross sections 
depend quadratically on the distributions. These cross sections
should therefore be highly sensitive to the form of the
parton distributions involved. A well-known example is the proposed
relation \cite{Brodsky,Ryskin-unpolarized} between the
unpolarized gluon distribution $g(x,Q^2)$ of the proton
and the exclusive $J/\psi$ leptoproduction cross section.
As opposed to light meson production, $J/\psi$ production amplitudes
depend only negligibly on quark distributions, which makes the
$J/\psi$ a very useful probe of the gluon content of nucleons.

Similarly, the polarized gluon distribution $\Delta g(x,Q^2)$
could possibly be investigated by means of measuring photon-proton
polarization asymmetries in exclusive $J/\psi$ leptoproduction.
This was considered in the small Bjorken $x$ limit in
\cite{Ryskin-polarized,Ryskin-workshop}.
In the present paper, we discuss the general case where the
process probes nonforward twist-2 gluon distributions.

In section \ref{section:nonfwd} we write down the amplitude for
$\gamma^* + p \rightarrow J/\psi + p$ in terms of nonforward
parton distributions. We restrict here to the case of collinear
scattering and follow the treatment of
\cite{Radyushkin,Hoodbhoy,MPW97,Ji-PRD55}. In order to evaluate
polarization asymmetries one then needs the helicity amplitudes
for the subprocess $\gamma^* + g \rightarrow J/\psi + g$.
These are derived in section \ref{section:helicity}, using the
nonrelativistic approximation for the $J/\psi$ wave function.

Contrary to the results of \cite{Ryskin-polarized,Ryskin-workshop},
the polarization asymmetry in hard exclusive $J/\Psi$ production
turns out to be zero at the twist-2 level. In addition, we have
found that the spin-flip transition between hard photon and $J/\psi$
is similarly suppressed. It implies that hard exclusive $J/\psi$
production is not a good probe of either the polarized gluon
distribution or the nonforward gluon helicity flip distribution
of the nucleon \cite{JiHoodbhoy98}.

\section{Nonforward distributions \label{section:nonfwd}}

Below we discuss the amplitude for the process
\beq
  \gamma^*(q,\lambda) + p(P,S)
  \rightarrow J/\psi(K,\lambda') + p(P',S') \, .
\label{process}
\eeq
Our presentation follows closely that of
Refs.\ \cite{Radyushkin,Hoodbhoy,Ji-PRD55} as well as the calculation of hard
exlusive production of light mesons in \cite{MPW97}. Our notation is
the same or analogous to that in \cite{Ji-PRD55}.

In Eq.\ (\ref{process}) $q$ is the four-momentum of the hard photon with
virtuality $Q^2 = - q^2$ and polarization $\lambda$. $P$,$S$ and
$P^\prime$,$S^\prime$ denote momenta and spins of initial and final nucleons,
respectively, and $K$,$\lambda^\prime$ is the momentum and polarization of the
produced $J/\psi$.
We shall first consider the amplitude for the process
(\ref{process}) in the limit when $Q^2$ is large, but our results are
presumably also valid in the photoproduction limit ($Q^2 = 0$), where
a hard scale is still provided by the charm quark mass.

\begin{figure}[b]
\centerline{\psfig{figure=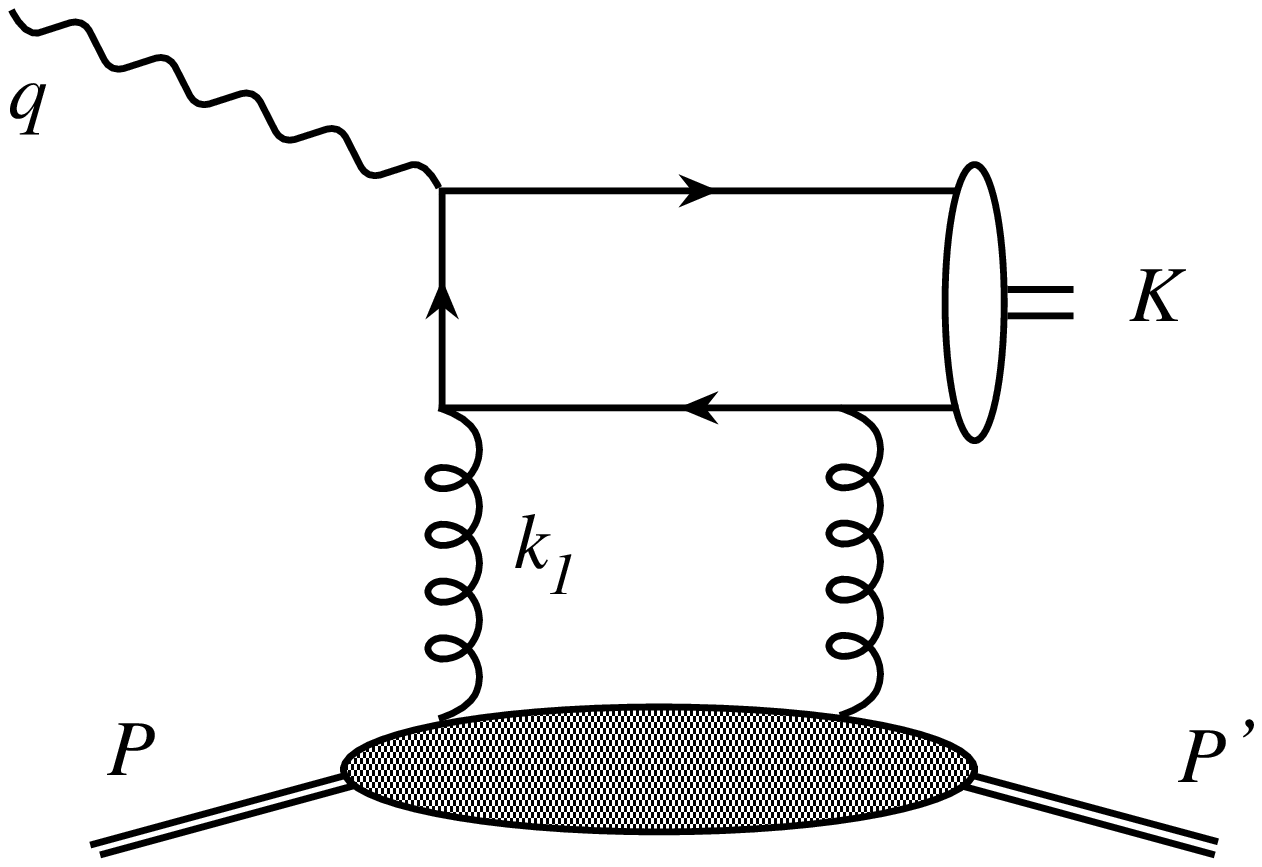,width=16cm}}
\caption[]{\sf One of 6 Feynman diagrams which contribute to the amplitude
(\ref{amplitude1}).} 
\label{diagram}
\end{figure}

In a general, covariant gauge, arbitrary number of gluons can be
exchanged between the charm quark loop and nucleon, all contributing
to the same order of the twist expansion. To avoid this complication,
in the following we choose the light-cone gauge for the gluon field.
In this case the dominant contribution comes from the exchange of
two gluons, as illustrated by the Feynman diagram on Fig.~\ref{diagram}.
The amplitude is first written as
\beq
  {\cal A}_{\lambda\lambda'SS'}
  = \int \frac{d^4 k_1}{(2\pi)^4}
     \, H^{\mu\nu}_{\lambda\lambda'}(q,K,k_1)
     \int d^4 z \, e^{i k_1 \cdot z}\, 
     \langle P',S' | A_\mu^a(-z/2) A_\nu^a(z/2) | P,S \rangle \, .
  \label{amplitude1}
\eeq
Here $H^{\mu\nu}_{\lambda\lambda'}(q,P,k)$
is the perturbative amplitude for the process
\beq
  \gamma^*(q,\lambda) + g(k_1)
  \rightarrow J/\psi(K,\lambda') + g(k_2) \, ,
\eeq
$k_1$ and $k_2$ are momenta of the incoming and outgoing gluons, respectively.

Calculating $H^{\mu\nu}_{\lambda\lambda'}$ to the leading twist
accuracy one can neglect the momentum transfer $t = (K-q)^2$ and
the nucleon mass $M_N$. The scattering amplitude then becomes
collinear, i.e., all momenta can be expressed as linear combinations
of two light-like vectors $P$ and $n$. Assuming that the space-like
component of the virtual photon momentum is directed along the
negative direction of the ${\hat z}$ axis, we can take 
$P=|{\bf P}|(1,0,0,1)$ and $n = (1,0,0,-1)/(2|{\bf P}|)$. The gluon
momenta $k_1$ and $k_2$ become collinear as well,
$k_i = (k_i\cdot n)P$. The amplitude can then be simplified to
\beqa
  {\cal A}_{\lambda\lambda'SS'}
  & = & \int_{-1}^{1} dx \int_{-\infty}^{\infty} \frac{d\lambda}{2\pi}
        \; e^{i\lambda x} \; H^{\mu\nu}_{\lambda\lambda'}(q,K,xP)
        \nonumber \\
  &   & \times
        \langle P',S' | A_\mu^a(-\lambda n/2) A_\nu^a(\lambda n/2)
        | P,S \rangle \, .
  \label{amplitude2}
\eeqa
Because of collinearity, the angular momentum conservation reduces to helicity
conservation. We can therefore select the case $S=S'$ in (\ref{amplitude2}) by
requiring that the photon and $J/\psi$ helicities are equal. In the light-cone
gauge $n \cdot A = 0$ the matrix element with equal proton spins is
parametrized by the unpolarized and polarized nonforward gluon distributions as
\beqa
  \lefteqn{\left. \langle P',S=\pm 1/2 | A_\mu^a(-z/2) A_\nu^a(z/2)
  | P,S=\pm 1/2 \rangle \right|_{z^2=0}}
  \nonumber \\   
  & = & \frac{1}{2} \int_{-1}^1 dx
        \frac{e^{-ix(P+P')\cdot z/2}}{(x-\xi+i\epsilon)(x+\xi-i\epsilon)}
        \nonumber \\
  &   & \times \sum_{\lambda_1\lambda_2} \delta_{\lambda_1\lambda_2}
        \epsilon_\mu(k_1,\lambda_1) \epsilon^*_\nu(k_2,\lambda_2)
        \; \left[
        G(x,\xi;\mu^2) \pm \lambda_1 \, \Delta G(x,\xi;\mu^2)
        \right] \, .
  \label{nonforward}
\eeqa 
Here 
\begin{equation}
\xi = \frac{(P - P^\prime) \cdot n}{(P + P^\prime) \cdot n}
\end{equation}
and
\begin{equation}
\epsilon_\mu(k,\lambda) = -\frac{1}{\sqrt{2}}(0,\lambda, i,0)
\label{polvect}
\end{equation}
are
polarization vectors for transverse gluons moving along the $z$ axis in the
positive direction. They coincide with the polarization vectors for gluons in
the light-cone gauge because $n \cdot \epsilon = 0$ in the collinear
kinematics.
The polarized and unpolarized nonforward gluon distributions
$G(x,\xi;\mu^2)$ and $\Delta G(x,\xi;\mu^2)$ are related to the
usual gluon distributions $g(x,\mu^2)$ and $\Delta g(x,\mu^2)$
as
\beqa
 G(x,0;\mu^2)        & = & x\, g(x,\mu^2) \, , \\
 \Delta G(x,0;\mu^2) & = & x\, \Delta g(x,\mu^2) \, .
\eeqa 
Inserting (\ref{nonforward}) into (\ref{amplitude2}) we obtain
\beqa
  {\cal A}_{\lambda\lambda \pm\pm}
  & = & \frac{1}{2} \int_{-1}^1 dx 
        \frac{1}{(x-\xi+i\epsilon)(x+\xi-i\epsilon)} 
        \nonumber \\
  &   & \times \sum_{\lambda_1}
        \; A_{\lambda\lambda\lambda_1\lambda_1}
        \; \left[
        G(x,\xi;\mu^2) \pm \lambda_1 \, \Delta G(x,\xi;\mu^2)
        \right] \, ,
  \label{amplitude3}
\eeqa          
where
\beq
  A_{\lambda\lambda'\lambda_1\lambda_2}
  = \epsilon_\mu(k_1,\lambda_1)
    H^{\mu\nu}_{\lambda\lambda'}(q,K,k_1=xP)
    \epsilon^*_\nu(k_2,\lambda_2)
\eeq
is the helicity amplitude for the perturbative subprocess. 

Eq.\ (\ref{amplitude3}) shows that in order to access the polarized
distribution $\Delta G(x,\xi;\mu^2)$, the structure 
$\lambda \lambda_1 \delta_{\lambda_1\lambda_2} \delta_{\lambda\lambda'}$
must be present in $A_{\lambda\lambda'\lambda_1\lambda_2}$.

\section{Helicity amplitudes for the subprocess \label{section:helicity}}

Let us now study the helicity amplitudes
$A_{\lambda\lambda'\lambda_1\lambda_2}$. To make contact with results existing
already in the literature \cite{Korner,BergerJones} we have chosen to consider
the general case i.e., 
retaining the full kinematics.
Using the standard nonrelativistic approximation of the $J/\psi$
wavefunction \cite{BergerJones}, we write
\beqa
  H^{\mu\nu}_{\lambda\lambda'}(q,K,k_1)
  & = & e Q_c g_s^2 \frac{R_S(0)}{\sqrt{16\pi M}}
        \frac{T_R}{\sqrt{N_C}}
        \; \epsilon_\alpha(q,\lambda)
        \, \epsilon^*_{\alpha'}(K,\lambda')
        \nonumber \\
  &   & \times \left\{
        {\rm Tr} \left[ \gamma^\alpha S_F(K/2-q) \gamma^{\mu}
                        S_F(-K/2-k_2) \gamma^{\nu} \gamma^{\alpha'}
                        (\slas{K} + M) \right]
        \right. \nonumber \\
  &   & \left. \mbox{}
        + \dots \right\}
  \label{charmtrace}
\eeqa
where $e$ is the electromagnetic and $g_s$ the strong coupling constant,
$Q_c=2/3$, $R_S(0)$ is the radial wavefunction of the $J/\psi$ at the
origin, $M=2m_c$ is the $J/\psi$ mass,
$S_F(k) \equiv (\slas{k} - m_c)^{-1}$, and the dots stand for five
other permutations of the gauge bosons on the charm quark line.
The term shown explicitly in
(\ref{charmtrace}) corresponds to the diagram shown in Fig.~\ref{diagram}.

It is convenient to express the polarization vectors in terms of
four-momenta of the problem. Clearly, it is not possible if the
scattering is collinear, but our expressions have the correct
limit as $\theta \rightarrow 0$.
We work in the Gottfried--Jackson frame, i.e.\ the $J/\psi$ rest frame
where the $J/\psi$ spin quantization direction is determined by the
virtual photon three-momentum. The $y$ axis is chosen in the direction
of ${\bf P} \times {\bf q}$. In this frame the polarization vectors are
\beqa
  \epsilon^\mu(q,\lambda=0) 
    & = & \frac{s+u}{T\sqrt{Q^2}}
          \left( q^\mu + \frac{2Q^2}{s+u} K^\mu \right) \, , \\
  \epsilon^{*\mu}(K,\lambda'=0) 
    & = & \frac{s+u}{MT}
          \left( -K^\mu + \frac{2M^2}{s+u} q^\mu \right) \, , \\
  \epsilon^\mu(q,\lambda=\pm 1)
    & = & N^{-1} \left[ \lambda R^\mu(\gamma^*) - iI^\mu \right] \, , \\  
  \epsilon^{*\mu}(K,\lambda'=\pm 1)
    & = & N^{-1} \left[ \lambda' R^\mu(\gamma^*) + iI^\mu \right] \, , \\  
  \epsilon^\mu(k_1,\lambda_1=\pm 1)
    & = & N^{-1} \left[ \lambda_1 R^\mu(g_1) - iI^\mu \right] \, , \\  
  \epsilon^{*\mu}(k_2,\lambda_2=\pm 1)
    & = & N^{-1} \left[ \lambda_2 R^\mu(g_2) + iI^\mu \right] \, ,
\eeqa
where
\beqa
  T^2 & = & (s+u)^2 + 4 M^2 Q^2 \, , \\
  N   & = & -\sqrt{2t(M^2 Q^2 + su)} \, , \\
  R^\mu(\gamma^*)
    & = & \frac{s+u}{T}
          \left[ \left( s+u + \frac{4M^2Q^2}{s+u} \right) k_1^\mu
          - \left( s+Q^2 - \frac{2Q^2(u-M^2)} {s+u} \right) K^\mu
          \right. \nonumber \\
    &   & \left. \mbox{}
          + \left( u-M^2 + \frac{2M^2(s+Q^2)} {s+u} \right) q^\mu
          \right] \, , \\
  R^\mu(g_1)
    & = & \left( s+u + \frac{2M^2(s+Q^2)}{u-M^2} \right) k_1^\mu
          + \left( s+Q^2 \right) K^\mu
          + \left( u-M^2 \right) q^\mu \, , \\
  R^\mu(g_2)
    & = & \left( s+u + \frac{2M^2(u+Q^2)}{s-M^2} \right) k_2^\mu
          - \left( u+Q^2 \right) K^\mu
          - \left( s-M^2 \right) q^\mu \, , \\
  I^\mu
    & = & 2 \epsilon^{\mu\nu\rho\sigma} K_\nu k_{1\rho} q_\sigma
          \, ,
\eeqa
and $s,t,u$ are the usual Mandelstam variables for the subprocess.
Writing
\beqa
  A_{\lambda \lambda' \lambda_1 \lambda_2}
    & = & \sqrt{\frac{\alpha_{\rm em}}{27 M}} \,
          \frac{32\pi\alpha_{\rm s} R_S(0)
          \; a_{\lambda \lambda' \lambda_1 \lambda_2}}{
          (s-M^2) (t-M^2-Q^2) (u-M^2) \, T^2}
\eeqa
we find the following helicity dependence:
\beqa
  a_{0 0 \lambda_1 \lambda_2}
  & = & \delta_{\lambda_1 \lambda_2} \cdot 2 \sqrt{Q^2}
        \, \left( M^2 Q^2 + su \right)
        \, \left[ \left(M^2+Q^2\right)^2 - t(t+4M^2) \right]
        \nonumber \\
  &   & \mbox{}
        + \delta_{\lambda_1, -\lambda_2} \cdot 2 \sqrt{Q^2} \, t M^2 T^2
        \, , \label{heli-structure1} \\
  a_{\lambda 0 \lambda_1 \lambda_2}
  & = & \delta_{\lambda_1 \lambda_2} \cdot N
        \, \left[ \lambda_1 T + \lambda (s-u) \right]
        \, \left[ (M^2+Q^2)^2 + t(Q^2-M^2) \right]
        \, ,\label{heli-structure3}\\
  a_{0 \lambda' \lambda_1 \lambda_2}
  & = & \delta_{\lambda_1 \lambda_2} \cdot N
        \, \left[ \lambda_1 T + \lambda' (u-s) \right]
        \cdot 2 t M \sqrt{Q^2}  \, , \label{heli-structure2}\\
  a_{\lambda \lambda' \lambda_1 \lambda_2}
  & = & \delta_{\lambda_1 \lambda_2} \delta_{\lambda \lambda'}
        \cdot 2 M (M^2 Q^2 + su) \left[-(M^2+Q^2)^2 + t(M^2-3Q^2) \right]
        \nonumber \\
  &   & \mbox{} 
        + \delta_{\lambda_1 \lambda_2} \delta_{\lambda, -\lambda'}
        \cdot tM(t-M^2-Q^2) \nonumber \\
  &   & \hspace{28mm} \times \, 
        \left[ \lambda_1 \lambda (s-u) T + s^2 + u^2 + 2M^2 Q^2 \right]
        \nonumber \\
  &   & \mbox{} 
        + \delta_{\lambda_1, -\lambda_2} \delta_{\lambda \lambda'}
        \cdot tMT^2 \, [\lambda_1 \lambda T - (s+u)]
        \, .      
  \label{heli-structure4}
\eeqa
In Eqs.\ (\ref{heli-structure1}-\ref{heli-structure4}) the indices
$\lambda,\lambda'$ refer to transverse polarization of the photon
and $J/\psi$, while longitudinal polarization is denoted by an
index "0".

In the limit $s \gg M^2 \sim Q^2 \gg |t|$ our helicity amplitudes
agree with Eq.\ (31) of \cite{Korner}. (Ref.\ \cite{Korner} also presents
an exact expression for the helicity amplitudes, but it is given in the
parton-level recoil frame rather than in the Gottfried--Jackson frame
and therefore cannot be directly compared with our result at $t\neq 0$.)
Our amplitudes are also valid at $Q^2=0$, in which case the charm
quark mass provides a large momentum scale, and reproduce the
unpolarized cross section of \cite{BergerJones}.

\section{Discussion}

First, from Eqs.\ (\ref{heli-structure1}-\ref{heli-structure4})
one recovers immediately the well known results for the leading
parts of the amplitudes $a_{\lambda \lambda' \lambda_1 \lambda_2}$
and $a_{0 0 \lambda_1 \lambda_2}$. They lead to helicity-conserving
$J/\psi$ production amplitudes proportional to nucleon unpolarized
gluon distributions \cite{Brodsky,Ryskin-unpolarized}. Moreover,
from (\ref{heli-structure4}) one finds that the structure
$\lambda \lambda_1 \delta_{\lambda_1\lambda_2} \delta_{\lambda\lambda'}$
is absent. We thus predict photon-proton polarization asymmetries
to vanish in the collinear limit relevant for Eq.\ (\ref{amplitude2}).

Let us briefly discuss the role of the helicity-flip structures
$\delta_{\lambda_1\lambda_2} \delta_{\lambda,-\lambda'}$ and
$\delta_{\lambda_1,-\lambda_2} \delta_{\lambda\lambda'}$ in
Eq.\ (\ref{heli-structure4}), which are potentially significant
for $J/\psi$ polarization studies and for the study of
gluon helicity-flip nonforward distributions \cite{JiHoodbhoy98},
respectively. In the Bjorken limit ($s \sim Q^2 \gg M^2, |t|$)
the helicity-flip terms are suppressed by a factor of $O(t/Q^2)$
with respect to the helicity-conserving 
$\delta_{\lambda_1\lambda_2} \delta_{\lambda\lambda'}$ term.
In the photoproduction limit, the situation depends on whether
the subprocess occurs close to the threshold ($s-M^2 \ll M^2$)
or far from the threshold ($s \gg M^2$). Close to the threshold
the ratio of the helicity-flip to the helicity-conserving term
becomes $t/(s-M^2+t)$. Far from the threshold the 
$\delta_{\lambda_1\lambda_2} \delta_{\lambda,-\lambda'}$ term
is suppressed by a factor $t/M^2$ and the 
$\delta_{\lambda_1,-\lambda_2} \delta_{\lambda\lambda'}$ term
by a factor $t(t-M^2)^2/(M^2 s^2)$.

Hence the $J/\psi$ photo- or leptoproduction process is not
a good probe of the gluon helicity-flip distribution,
except possibly close to the subprocess threshold.
A helicity flip between the photon and the $J/\psi$ should
become significant in photoproduction at $|t| \gsim M^2$.

In order to cross-check our results we have calculated the amplitude
(\ref{amplitude2}) directly in the collinear limit using the polarization
vectors (\ref{polvect}). We have found the contribution from the
unpolarized gluon distribution which agrees exactly with
Refs.\ \cite{Brodsky,Ryskin-unpolarized}. As far as the
polarized gluon 
distribution is concerned, we have found that the contribution of each
individual diagram has exactly the same magnitude as in the unpolarized case,
but half of them have opposite signs due to asymmetry with respect to exchange
of $\mu \leftrightarrow \nu$ in the part of the matrix element
(\ref{nonforward}) which is proportional to $\Delta g(x,\xi,\mu^2)$. As a 
consequence, the total contribution is zero.
Our result thus corrects earlier work
\cite{Ryskin-polarized,Ryskin-workshop} where there was a mistake,
as confirmed by the author \cite{Ryskin-private}.

In summary, we have discussed the calculation of exclusive 
$J/\psi$ production off nucleons. We have found that the contribution of
twist-2 polarized gluon distribution vanishes in the hard virtual photon limit.
It will thus be difficult to
determine polarized gluon distribution from polarization asymmetry in hard
exclusive $J/\psi$ production. We note, however, that the present calculation
has been performed neglecting the relative velocity $v$ of
charmed quarks in the $J/\psi$ bound state. It is possible that relaxing this
approximation, like e.g., in \cite{Hoodbhoy}, will lead to non-vanishing
polarization asymmetry as well as a non-zero coupling to gluon helicity-flip
distribution at the leading twist level, but at the subleading order in the
$v/c$ expansion. 

\bigskip

{\bf Acknowledgement.} We wish to thank M.~G.~Ryskin for useful discussions.

\end{document}